# Spatially Encoded Pseudo-pure States for NMR Quantum Information Processing

Yehuda Sharf[1], Timothy F. Havel[2], and David G. Cory[1,*]

[1]*Dept. of Nuclear Engineering, Massachusetts Institute of Technology, Cambridge, MA 02139, USA*
[2]*BCMP, Harvard Medical School, Boston, MA 02115, USA*

## ABSTRACT

Quantum information processing by liquid-state NMR spectroscopy uses pseudo-pure states to mimic the evolution and observations on true pure states. A new method of preparing pseudo-pure states is described, which involves the selection of spatially labeled states of an ancilla spin with which the spin system of interest is correlated. This permits a general procedure to be given for the preparation of pseudo-pure states on any number of spins, subject to the limitations imposed by the loss of signal from the selected subensemble. The preparation of a single pseudo-pure state is demonstrated by carbon and proton NMR of labeled alanine. With a judicious choice of magnetic field gradients, the method further allows encoding of up to $2^N$ pseudo-pure states in independent spatial modes in an N+1 spin system. Fast encoding and decoding schemes are demonstrated for the preparation of four such spatially labeled pseudo-pure states.

---

[*] To whom correspondence should be addressed at dcory@mit.edu.

# I. INTRODUCTION

Quantum information processing requires that the system be placed in a "fiducial" state, relative to which information can be stored in it [1]. This is usually taken to be a pure state $|0\rangle$ of an array of *N* two-state quantum systems (qubits), whose basis states then correspond to a binary encoding of the integers from 0 to $2^N - 1$. Such a pure state encoding allows the system to be placed in a coherent superposition over all the basis states, so that by the linearity of quantum mechanics unitary operations can be applied to all these states in parallel. In many of the proposed implementations of quantum computers, the preparation of a known pure state nevertheless poses a substantial experimental challenge. In the case of nuclear spin systems, for example, it requires either Stern Gerlach apparatus [2], cooling the system to micro-Kelvin temperatures [3], optical pumping [4] or dynamic nuclear polarization [5].

An alternative to true pure states is a special kind of mixed state known as a pseudo-pure state [6, 7]. Such a state is characterized by a density matrix with a single non-degenerate eigenvalue, whose corresponding eigenvector transforms identically to the state vector of the same spin system in a pure state. Unlike true pure states, a pseudo-pure state is necessarily the state of an ensemble, but it is straightforward to show that the expectation value of a traceless observable over a pseudo-pure ensemble is proportional to the expectation value of the same observable in the corresponding pure state. It follows that the dynamics of the observables on pseudo-pure states is equivalent to those of pure states. They have played an essential role in all demonstrations of quantum information processing by liquid-state NMR to date [8, 9, 10, 11].

Because the density matrix of a pseudo-pure state does not have the same eigenvalues as those of the thermal equilibrium state, the preparation of pseudo-pure states necessarily involves an incoherent averaging process. To date, five different kinds of pseudo-pure states have been proposed [12]:

1) A relative pseudo-pure state, obtained by taking the partial trace over an auxiliary spin system with which the system of interest is correlated; as shown in [13], this is done in NMR by decoupling the auxiliary system during acquisition (observation).
2) A conditional pseudo-pure state, obtained by filtering the system conditional on the state of an auxiliary spin system with which it is correlated (this is called a logically labeled effective pure state in [8] ); in NMR, filtering is done by selecting components from the multiplets due to coupling with the auxiliary system.

3) A temporally average pseudo-pure state, obtained by averaging the results of different experiments over multiple mixed states whose mean density matrix is pseudo-pure [14]; this is similar to the use of phase cycling in NMR.
4) A spatially average pseudo-pure state, which is obtained by creating a spatial distribution of states across the ensemble whose mean density matrix is pseudo-pure; in NMR, this is conveniently done by means of magnetic field gradients [10].
5) A spatially conditional pseudo-pure state [15], wherein the spatial average yields a conditional pseudo-pure state as in (2) above.

Here we describe a new method of preparing spatially conditional pseudo-pure states by NMR. This requires only a single ancilla for any number of pseudo-pure spins, and has the advantage over previous methods of making it straightforward to give a closed procedure, which can in principle generate a pseudo-pure state on any number of spins. The method involves creating correlations between the spin system of interest and an ancilla spin, whose phase in turn is correlated with its spatial position [16]. By selectively refocussing the ancilla spin, it is possible to pick out a pseudo-pure basis state for the system of interest, conditional on the state of the ancilla. It is further possible to create a classical ensemble containing all $2^N$ pseudo-pure states, each with a different spatial labeling, and to read out the results obtained from identical computations on all of these states in a single experiment.

As with all of the above forms of pseudo-pure states, the preparation of a spatially conditional pseudo-pure state from the corresponding high-temperature equilibrium state necessarily entails a loss of signal that is exponential in the number of spins involved [17]. Nevertheless, in the procedures given here it is at least easy to keep track of all the magnetization components.

## II. THE PROJECTION OPERATOR

The method relies on a conditional phase shift, $D_{k_1 k_2}^{a|i}$, of the ancilla following the application of the fundamental sequence: a controlled-NOT (c-NOT) gate, a selective magnetic field gradient pulse, another c-NOT gate, and a rephasing selective gradient pulse (see Fig. 1). With the magnetic field gradients applied along the $z$-axis the corresponding propagator in geometric algebra notation [18] is

$$D_{k_1 k_2}^{a|i} \equiv G_{k_2}^a \left( \sigma_x^a E_-^i + E_+^i \right) G_{k_1}^a \left( \sigma_x^a E_-^i + E_+^i \right) \tag{1}$$

where

$$G_k^a \equiv e^{-\frac{i}{2}\Delta z \sum k^j \sigma_z^j} e^{-i\frac{\pi}{2}\sigma_y^a} e^{\frac{i}{2}\Delta z \sum k^j \sigma_z^j} e^{i\frac{\pi}{2}\sigma_y^a} = e^{-i\Delta z k^a \sigma_z^a}. \quad (2)$$

Here, $\sigma_\eta^a$ are Pauli matrices and $E_\pm^i = \frac{1}{2}(1_2 \pm \sigma_z^i)$ are idempotent operators associated with the population of the $i-th$ spin state. The superscript $a$ indicates that magnetic field gradients are applied to selectively shift the phase the ancilla. Gradient selectivity ($G_k^a$) is accomplished using a combination of two selective refocusing RF $\pi$ pulses and two magnetic field gradient pulses with alternate polarities (Fig. 1). The wave number $k^j(\Delta t) \equiv \gamma_j \int_0^{\Delta t} \frac{\partial B_z}{\partial z} dt$ is determined by the pulse shape of the magnetic field gradient $\frac{\partial B_z}{\partial z}$, and by the gyromagnetic ratio $\gamma_j$ of the $j$-th spin.

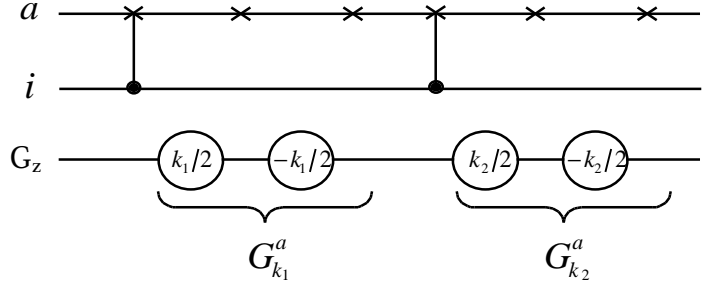

**Figure 1:** A circuit for the conditional phase shift operator $D_{k_1 k_2}^{ai}$. Single-spin (NOT) and two-spin (c-NOT) operations affect only the ancilla (top) and the $i$-th qubit (middle). The bottom line shows the relative amplitude and polarity of the applied magnetic field gradient pulses.

The effect of the RF pulses and magnetic field gradients can be described using Average Hamiltonian Theory (AHT) [19]. The propagator in Eq. (1) can be written as

$$\begin{aligned} D_{k_1 k_2}^{ai} &= e^{-i\Delta z k_2^a \sigma_z^a}\left(\sigma_x^a E_-^i + E_+^i\right) e^{-i\Delta z k_1^a \sigma_z^a}\left(\sigma_x^a E_-^i + E_+^i\right) \\ &= e^{-i\Delta z\left(k_2^a - k_1^a\right)\sigma_z^a} E_-^i + e^{-i\Delta z\left(k_2^a + k_1^a\right)\sigma_z^a} E_+^i \\ &= e^{-i\Delta z \sigma_z^a \left[\left(k_2^a - k_1^a\right)E_-^i + \left(k_2^a + k_1^a\right)E_+^i\right]} \end{aligned} \quad (3)$$

yielding an averaged Hamiltonian of

$$\bar{H} = \sigma_z^a \Delta z \left( (k_2 - k_1) E_-^i + (k_1 + k_2) E_+^i \right). \tag{4}$$

This spatially varying gradient Hamiltonian commutes with $\sigma_z^a$ and therefore does not affect the z-component of the ancilla magnetization. It introduces a spatially varying phase which when averaged over the ensemble renders any transverse components of the ancilla unobservable. The conditional phase shift winds the transverse components of the ancilla's magnetization into a spatial helix along the z-axis. The average magnetization is thus zero (neglecting edge effects), but can be returned to its original uniform phase by reversing the conditional phase shift. Molecular diffusion limits the possibility of refocusing; the conditional phase shift correlates the magnetization phase with location, whereas molecular diffusion blurs this correlation with an accompanied loss of information. Therefore in the presence of molecular diffusion, which is always present in liquid-state NMR, the Hamiltonian in Eq. (4) may be regarded as decoherent.

One of the terms in Eq. (4) may be made to vanish (e.g. $k_1 = \pm k_2 = k$), so that the effective Hamiltonian reduces to

$$H_{k_1 = \mp k_2 = k} = 2k \Delta z \sigma_z^a E_\pm^i. \tag{5}$$

We observe that if the ancilla is correlated with the *i-th* spin in the state $E_+^i$ it will remain unaffected by the Hamiltonian $2k\Delta z \sigma_z^a E_-^i$. Similarly if it is correlated with the state $E_-^i$ it will remain unaffected by the Hamiltonian $2k\Delta z \sigma_z^a E_+^i$. Otherwise, its transverse magnetization will acquire a spatial phase of $2k\Delta z$ and will become undetectable. It is easy to see that reintroducing the same Hamiltonian, i.e. by reapplying the conditional phase shift with the same gradient setting, will tighten the grating but will not alter the observable magnetization. In the absence of molecular diffusion it is also possible to rewind the spatial grating, simply by reverting the gradients, and to retain the original state. Allowing sufficient time (following the conditional phase shift) for molecular diffusion to inhomogeneously destroy the grating leads to irreversible loss of all information stored in the non-zero spatial encoded phase. Thus with a proper selection of magnetic field gradients the conditional phase shift $D_{k_2 = \pm k_1}^{a|i}$ performs a projection of an arbitrary state into a subspace that spans the ancilla and one of the *i-th* spin eigenstates. It is projective in the sense that, unlike other pseudo-pure preparation methods, it selects the magnetization component in a chosen subspace regardless of the initial state of the system and therefore can be applied repeatedly without changing the result.

## III. APPLICATION TO THE EQUILIBRIUM STATE

The scheme presented above (Fig. 1) implements the projection Hamiltonian of Eq. (5) without any assumption on the initial state of the spin system. If, however, the initial state is known, e.g. thermal equilibrium, the scheme can be further simplified. In the following we will discuss the extension of the projection propagator to an arbitrarily large number of qubits, and we will demonstrate how it can be utilized for pseudo-pure state preparation from a spin system initially in thermal equilibrium.

The general preparation scheme is described in Fig. 2. It consists of two parts: a) The creation of correlations between the ancilla and the data spins; b) Application of a projective *k*-space labeling to the desired subspace. One can look at this as a two step process, each involving the application of *N c-NOT* gates. The first step ensures maximum transfer of polarization, whereas the actual encoding is done in the second step, as described in the next two subsections.

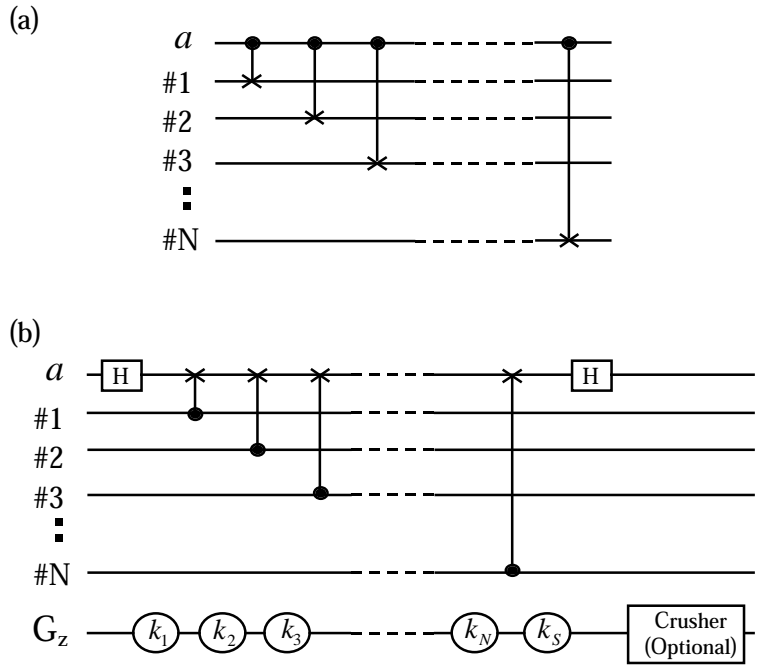

**Figure 2:** Two-step procedure for preparing spatially encoded pseudo-pure states for a spin system initially in thermal equilibrium. (a) To allow a maximum transfer of polarization the magnetization of the data spins is correlated with that of the ancilla. (b) The actual k-space encoding is carried out via a sequence of *N c-NOT* gates and selective magnetic field gradients (see text). The symbol •—× corresponds to a controlled-NOT, i.e. a gate which flips the target bit (•) if the control bit (×) is in the state $|1\rangle$, and the symbol H stands for a Hadamard gate.

## A. Correlating the ancilla with the data spins

The deviation part of the state of a spin system in thermal equilibrium is

$$\rho_{eq} = \sigma_z^a + \sum \frac{\gamma_i}{\gamma_a} \sigma_z^i \qquad (6)$$

where common factors have been omitted. Each application of a $c-NOT_{ai}$ gate (i.e. flipping the $i$-th qubit conditional on the state of the ancilla) correlates the magnetization of the $i$-th data spin with the ancilla, and the state after $N$ successive applications (Fig. 2(a)) is

$$\rho_0 = \sigma_z^a \left( 1 + \sum_{i=1}^{N} \frac{\gamma_i}{\gamma_a} \sigma_z^i \right). \qquad (7)$$

Following $N$ c-NOTs the state of the ancilla can be factored out indicating its correlation with all the data spins. The state can also be conveniently expressed in the standard computational basis as [20]

$$\rho_0 = \sigma_z^a \sum_{b_1 b_2 \ldots b_N = 0}^{1} (1 + \varepsilon_{b_1 b_2 \ldots b_N}) |b_1 b_2 \ldots b_N\rangle\langle b_1 b_2 \ldots b_N| \\ = \sigma_z^a \sum_{\alpha=0}^{2^N - 1} (1 + \varepsilon_\alpha) |\alpha\rangle\langle\alpha| \qquad (8)$$

where $|\alpha\rangle\langle\alpha|$ stands for the binary representation of the $2^N$ basis functions set for the data qubits, and $\varepsilon_{b_1 b_2 \ldots b_N} \equiv \sum_{i=1}^{N} \frac{\gamma_i}{\gamma_a} (-1)^{b_i}$ is the relative energy associated with the eigenstate $|b_1 b_2 \ldots b_N\rangle$.

## B. Pseudo-pure state projection

The purpose of the k-space encoding procedure described in Fig. 2(b) is to project the state $\rho_0$ into a subspace corresponding to a pre-selected eigenstate $|\alpha\rangle$. It is an extension of the conditional phase shift described above. The first Hadamard gate tilts the magnetization of the ancilla to the transverse plane ($\sigma_x^a$). Since the state of the data spins is diagonal, the conditional spatial phasing operator needed for each data qubit can be simplified to

$$D_k^{a|i} \equiv \left(\sigma_x^a E_-^i + E_+^i\right) e^{-\frac{i}{2}k\Delta z \sum_j \sigma_z^j}. \tag{9}$$

In fact, the reduced operator is effectively equivalent to the phase shift operator in Eq. (3) without the rephasing gradient (i.e. $k_2 = 0$). The latter is applied instead at the end of the procedure (Fig. 2(b)). The successive application of N phasing operators ($D_{k_n}^{a|n}$; $n = 1\ldots N$) yields

$$\begin{aligned}
D_{k_N}^{a|N}\ldots D_{k_2}^{a|2} D_{k_1}^{a|1} &\left[\sigma_x^a \sum_{\alpha=0}^{2^N-1}(1+\varepsilon_\alpha)|\alpha\rangle\langle\alpha|\right]\left(D_{k_1}^{a|1}\right)^{-1}\left(D_{k_2}^{a|2}\right)^{-1}\ldots\left(D_{k_N}^{a|N}\right)^{-1} \\
&= \sum_{\alpha=0}^{2^N-1} e^{-ik_\alpha \Delta z \sigma_z^a} \sigma_x^a e^{ik_\alpha \Delta z \sigma_z^a} \varepsilon_\alpha |\alpha\rangle\langle\alpha|
\end{aligned} \tag{10}$$

where in the last step the properties $E_\pm^i E_\pm^i = E_\pm^i$, $E_\pm^i E_\mp^i = 0$ have been used. The k-space labels

$$k_\alpha = \sum_{n=1}^{N} k_n P_n(\alpha) \tag{11}$$

are determined by the strength and duration of the gradients ($k_n$) applied at each step, as well as by the parity $P_n(\alpha) \equiv \prod_{i=n}^{N}(-1)^{b_i}$ of the sum of the last $N-n+1$ qubits in the binary representation of the state $\alpha$.

The magnetic field gradient pulse $k_s$ applied at the end of the encoding procedure (Fig. 2(b)) adds a fixed phase to the ancilla. Therefore, in order to rephase terms labeled by $k_{\alpha'}$ it has to satisfy the condition

$$k_s = -k_{\alpha'}. \tag{12}$$

Since the k-labels in Eq. (11) may in general be degenerate this condition may apply to more than one subspace.

## IV. K-SPACE ENCODING OF A SINGLE PSEUDO-PURE STATE OF N DATA SPINS

The general scheme in Fig. 2(b) can be used to prepare a system of N data spins in any pseudo-pure basis state. Consider the case of magnetic field gradient pulses of fixed strength and duration. This implies that $k_n = k_0$ for $n = 1...N$, and according to Eq. (11) the k-labels are

$$k_\alpha = k_0 \sum_{n=1}^{N} P_n(\alpha). \qquad (13)$$

The wave number $k_\alpha$ reaches its absolute maximum of $Nk_0$ at the ground state ($|\alpha = 0\rangle$) where $P_n(0) = 1$ for all $n$. Thus, in order to select this particular pseudo-pure state the selection gradient must satisfy, according to the Eq. (12), the condition $k_s = -Nk_0$. Here, the negative polarity of the last gradient is obtained by reversing the electric current to the gradient coils.

It is convenient to follow the evolution of the spatial phase acquired by the ancilla in the computational basis $|\alpha\rangle$. This phase can be listed in a table where each column, starting from the left, represents the outcome following an encoding step, i.e. a selective magnetic field gradient or a conditional rotation, and the final k-space encoding is given in the right column. An example for the spatial encoding of the pseudo-pure state $|\alpha = 0\rangle$, the case of three data qubits ($N = 3$) is illustrated in Table I. At the end of the encoding procedure a gradient of $k_s = -3k_0$ is applied (right column) so that only the $|000\rangle$ subspace is rephased, as required.

Likewise the data spins can be prepared in any arbitrary pseudo-pure state $|\alpha'\rangle$ by alternating the polarities of the applied gradient pulses. It follows from Eq. (11) that the state $|\alpha'\rangle$ can be prepared by choosing $k_n$ so that the condition

$$k_n = k_0 P_n(\alpha') \qquad (14)$$

is satisfied for all $n = 1...N$, whereas as before the selection gradient is $k_s = -k_{\alpha'} = -Nk_0$. For example, to prepare the state $|\alpha' = 2\rangle = |010\rangle$ in a system of three data qubits: $p_1(|010\rangle) = (1)(-1)(1) = -1$, $p_2 = -1$ and $p_3 = 1$.

TABLE I. The acquired spatial phase of the ancilla (in units of $\frac{1}{2}\Delta z$) during encoding of a single pseudo-pure state.

| Subspace $\|b_1 b_2 b_3\rangle$ | $k_1 = k_0$ | $CNOT_{1a}$ | $k_2 = k_0$ | $CNOT_{2a}$ | $k_3 = k_0$ | $CNOT_{3a}$ | $k_s = -3k_0$ |
|---|---|---|---|---|---|---|---|
| $\|000\rangle$ | $k_0$ | $k_0$ | $2k_0$ | $2k_0$ | $3k_0$ | $3k_0$ | $0$ |
| $\|001\rangle$ | $k_0$ | $k_0$ | $2k_0$ | $2k_0$ | $3k_0$ | $-3k_0$ | $-6k_0$ |
| $\|010\rangle$ | $k_0$ | $k_0$ | $2k_0$ | $-2k_0$ | $-k_0$ | $-k_0$ | $-4k_0$ |
| $\|011\rangle$ | $k_0$ | $k_0$ | $2k_0$ | $-2k_0$ | $-k_0$ | $k_0$ | $-2k_0$ |
| $\|100\rangle$ | $k_0$ | $-k_0$ | $0$ | $0$ | $k_0$ | $k_0$ | $-2k_0$ |
| $\|101\rangle$ | $k_0$ | $-k_0$ | $0$ | $0$ | $k_0$ | $-k_0$ | $-4k_0$ |
| $\|110\rangle$ | $k_0$ | $-k_0$ | $0$ | $0$ | $k_0$ | $k_0$ | $-2k_0$ |
| $\|111\rangle$ | $k_0$ | $-k_0$ | $0$ | $0$ | $k_0$ | $-k_0$ | $-4k_0$ |

The application of final gradient pulse according to condition Eq. (12) yields the state

$$\sigma_x^a \varepsilon_\alpha |\alpha'\rangle\langle\alpha'| + \sum_{\substack{\alpha=0 \\ \alpha \neq \alpha'}}^{2^N-1} e^{-i(k_\alpha + k_S)\Delta z \sigma_z^a} \sigma_x^a e^{i(k_\alpha + k_S)\Delta z \sigma_z^a} \varepsilon_\alpha |\alpha\rangle\langle\alpha|. \quad (15)$$

Our choice of k-space labels based on a single maximum ensures that $k_\alpha + k_S \neq 0$ for all terms in the sum. These terms will vanish on averaging over the *z*-coordinate of the ensemble, leaving only the subspace $|\alpha'\rangle$ observable.

Finally, the second Hadamard gate at the end of the encoding procedure (Fig. 2(b)) tilts the magnetization of the ancilla back to the *z*-axis, leading to the desired spatially conditional pseudo-pure state

$$\sigma_z^a \varepsilon_{\alpha'} |\alpha'\rangle\langle\alpha'|. \quad (16)$$

## V. EXPERIMENTAL DEMONSTRATION OF A SINGLE PSEUDO-PURE STATE

Implementation is greatly simplified by the fact that the magnetization of the data spins remains along the z-axis throughout the preparation procedure. Magnetic field gradients have no effect on this longitudinal magnetization so that only the transverse components of the ancilla's magnetization are dephased. Diagonal terms (z-components) of the density matrix commute with the internal Hamiltonian, thereby reducing the requirement for elaborate methods of refocusing.

The scheme in Fig. 1 was implemented in liquid-state proton and carbon NMR of $^{13}C$ labeled alanine $\left(NH_3^+ - C^\alpha H(C^\beta H_3) - C' O_2'\right)$. Measurements were conducted on a Bruker AMX400 spectrometer (9.6 T) equipped with a 5 mm probe tuned to $^{13}C$ and $^1H$ frequencies of 100.61 MHz and 400.13 MHz, respectively. The probe was equipped with xyz-gradient coils capable of generating field gradients ranging from -60 to +60 G/cm.

With decoupling of the methyl protons, alanine exhibits a weakly coupled four-spin system. The $C^\alpha$ was chosen as the ancilla because of its well-resolved couplings with the other carbons ($J_{C^\alpha C'}$=35.1 Hz , $J_{C^\alpha C^\beta}$=54.2 Hz) and with the adjacent proton ($J_{C^\alpha H}$=143Hz), which were therefore chosen as the three data spins.

Logic gates were implemented using time delays and selective RF excitations. The latter included Gaussian shaped pulses for $\pi/2$ excitations and adiabatic pulses for inversions. For c-NOT gates ($N^{a|i}$) the following sequence was used [21]

$$\left(\frac{\pi}{2}\right)^a_{-x} - \left(\frac{\pi}{2}\right)^a_{-y} - \frac{1}{2J_{a,i}} - \left(\frac{\pi}{2}\right)^a_y, \tag{17}$$

where $\left(\frac{\pi}{2}\right)^a_\varphi$ are selective pulses about the $\varphi$ axis applied to the ancilla and $1/(2J_{a,i})$ is an evolution under the effective Hamiltonian $\frac{\pi}{2}J_{a,i}\sigma^a_z\sigma^i_z$ for a time $1/(2J_{a,i})$. Hadamard gates were realized by $(\pi/2)^a_{\pm y}$ rotations.

The preparation of the three data qubits in a single pseudo-pure state is demonstrated in Fig. 3. By skipping the first part of correlating the ancilla with the data spins (Fig. 2(a)), we choose to demonstrate the non-trivial part of the encoding process (Fig. 2(b)) on the equilibrium state of the ancilla alone: $\sigma^a_z$. The equilibrium magnetization of the data spins was effectively removed using selective $\pi/2$ RF pulses followed by a strong (crusher) magnetic field gradient. To illustrate the progress of k-space encoding, the spectrum of the ancilla obtained after each step in the encoding process is shown in Fig. 3. The exponential reduction in the number of peaks as a function of the number of steps, i.e. from 8 peaks to a single peak after N=3 applications of conditional phase shifts, demonstrates the efficiency of the method.

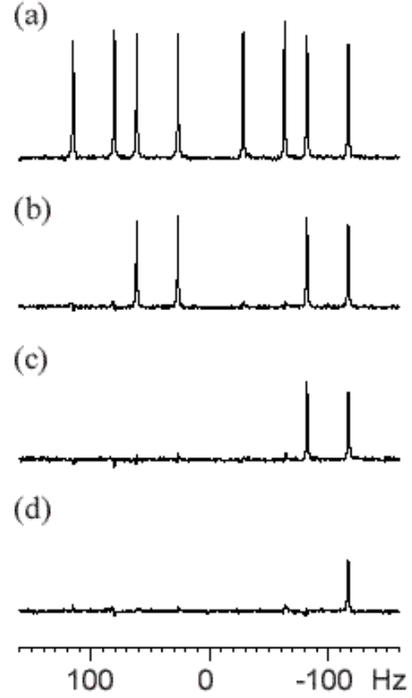

**Figure 3:** Preparation of a spatially conditional pseudo-pure state of three data spins initially in the state $\sigma_z^a$. In (a) the spectrum after applying a selective $(\pi/2)_y^a$ monitoring pulse exhibits 8 peaks due to J-couplings between the ancilla ($C^\alpha$) and the two carbons ($J_{C^\alpha C'} = 54.2$ Hz and $J_{C^\alpha C^\beta} = 35.1$ Hz), and with the adjacent proton ($J_{C^\alpha H} = 143$ Hz). To illustrate the progress of the encoding procedure (Fig. 2(b)), each encoding step was followed by a rephasing magnetization field gradient pulse and a selective $(\pi/2)_y^a$ RF monitoring pulse. Spectrum obtained: (b) after a single encoding step corresponding to the state $\sigma_x^a E_+^1$; (c) after two steps corresponding to the state $\sigma_x^a E_+^1 E_+^2$; and finally (d) after three steps corresponding to the three-spin pseudo-pure state $\sigma_x^a E_+^1 E_+^2 E_+^3$ (where the carbonyl $C'$, the proton, and $C^\beta$ were chosen as the first, second and third data qubits, respectively). The exponential decrease in the number of peaks demonstrates the efficiency of the method.

# VI. K-SPACE LABELING OF $2^N$ PSEUDO-PURE STATES

## A. Encoding

The preparation of a single pseudo-pure state required only gradient pulses of equal absolute strength, and criteria for the rephasing of a desired subspace were derived from the requirement for a single optimum. This insured the dephasing of all the subspaces but the desired one. Following the encoding procedure the final phase $k_\alpha + k_S$ of some of the terms in the sum (Eq.(15)) are degenerate (see Table I). However this redundancy was not important as long as their $k$-labels were non-zero.

This degeneracy can be lifted if the $k_n$ in Eq. (11) are allowed to take any value. The complete set of $2^N$ pseudo-pure states can be labeled, for example, with the following assignments

$$k_n = (-2)^{n-1} k_0. \tag{18}$$

Each application of the conditional phasing operator $D_{k_n}^{a|n}$ polarizes the ancila with different phase so that at the end of the procedure (Fig. 2(b)) all subspaces are uniquely labeled. This is illustrated for the case of 3 data spins in Table II.

TABLE II. The spatial phase of the ancilla acquired during multiple encoding of all eight pseudo-pure states provided by three data qubits.

| Subspace $\|b_1 b_2 b_3\rangle$ | $k_1 = k_0$ | $CNOT_{1a}$ | $k_2 = -2k_0$ | $CNOT_{2a}$ | $k_3 = 4k_0$ | $CNOT_{3a}$ |
|---|---|---|---|---|---|---|
| $\|000\rangle$ | $k_0$ | $k_0$ | $-k_0$ | $-k_0$ | $3k_0$ | $3k_0$ |
| $\|001\rangle$ | $k_0$ | $k_0$ | $-k_0$ | $-k_0$ | $3k_0$ | $-3k_0$ |
| $\|010\rangle$ | $k_0$ | $k_0$ | $-k_0$ | $k_0$ | $5k_0$ | $5k_0$ |
| $\|011\rangle$ | $k_0$ | $k_0$ | $-k_0$ | $k_0$ | $5k_0$ | $-5k_0$ |
| $\|100\rangle$ | $k_0$ | $-k_0$ | $-3k_0$ | $-3k_0$ | $k_0$ | $k_0$ |
| $\|101\rangle$ | $k_0$ | $-k_0$ | $-3k_0$ | $-3k_0$ | $k_0$ | $-k_0$ |
| $\|110\rangle$ | $k_0$ | $-k_0$ | $-3k_0$ | $3k_0$ | $7k_0$ | $7k_0$ |
| $\|111\rangle$ | $k_0$ | $-k_0$ | $-3k_0$ | $3k_0$ | $7k_0$ | $-7k_0$ |

The table shows that at the end of the process each subspace is associated with a different k-label. A general property of this encoding (Eq. (18)), which is also exhibited in the last column of Table II, is that wave numbers are discretely distributed. The $2^N$ k-labels range from $(1-2^N)k_0$ to $(2^N-1)k_0$ and with equal spacing

$$\Delta k = 2k_0, \qquad (19)$$

where adjacent k-labels correspond to states separated by a single spin flip.

Immediately after the application of the last conditional phase shift $D_{k_N}^{a|N}$ the magnetization of the ancilla lies in the transverse plane, and its phase is spatially encoded according to Eq. (11). It is advantageous to rotate the magnetization back to the *z*-direction because of the slower decay rate ($T_1$ vs. $T_2$), and also because the *z*-component commutes with the Zeeman term of the internal Hamiltonian as well as with the magnetic field gradients. However, a Hadamard rotation brings only the *x*-component of the magnetization back to the *z*-axis ( $H\sigma_x = \sigma_z$), leaving the *y*-component in the transverse plane ( $H\sigma_y = -\sigma_y$). Moreover, the Hadamard rotation destroys the distinguishability of positive and negative k-labels, i.e. following the rotation it is impossible to refocus the subspace labeled by $k_\alpha$ without refocusing the one labeled $-k_\alpha$. A way to solve the latter problem is to apply, prior to the Hadamard gate, an additional gradient pulse to add a constant phase shift of $k_s = 2^N k_0$ so that all wave numbers become positive. Finally, the y-component is removed using a strong (crusher) gradient applied immediately after the Hadamard gate (Fig. 2(b)) with the expense of half of the total magnetization. This procedure leads to the encoded state

$$\rho_e = \tfrac{1}{2}\sigma_z^a \sum_{\alpha=0}^{2^N-1} \cos((k_\alpha + k_s)\Delta z)\varepsilon_\alpha |\alpha\rangle\langle\alpha| \qquad (20)$$

### B. Decoding

At the end of the encoding procedure all subspaces are encoded with some spatial phase (Eq. (20)), so that a $\pi/2$ monitoring pulse applied to the ancilla yields no observed magnetization. Therefore the following decoding scheme is used

$$\left(\tfrac{\pi}{2}\right)_y^a - G_k^a \qquad (21)$$

where $(\pi/2)_y^a$ is an RF pulse applied selectively to the ancilla. Gradient selectivity ($G_k^a$) is accomplished using a combination of two selective refocusing RF pulses and magnetic field gradient pulses with alternate polarities. The selectivity of the magnetic field gradients prevents potential dephasing due to transverse magnetization of the other spins.

The effect of the decoding sequence (Eq. (21)) is to create a gradient echo for a selected subspace. Such an echo will occur only when the condition

$$k = -(k_\alpha + k_s) \qquad (22)$$

is satisfied. Thus (assuming the data spins are all along the z-axis), a weak readout gradient employed continuously during detection would give rise to gradient echoes at $t_n$ such that $k = \gamma_a \frac{\partial B}{\partial z} t_n = (2n+1)k_0$. To illustrate this effect the four subspaces associated with two data spins ($C'$ and $C^\beta$) were spatially encoded using the condition in Eq. (18). Starting with the state $\sigma_z^a$ the resultant encoded state was $\tfrac{1}{2}\sigma_z^a \sum_{\alpha=0}^{3} \cos(k_\alpha + k_S)|\alpha\rangle\langle\alpha|$ where $k_\alpha + k_S = 3k_0$, $5k_0$, $k_0$, and $7k_0$ correspond to the pseudo-pure states $\alpha = 0$ to 3, respectively. During encoding magnetic field gradient pulses of 2.5 gauss/cm and total duration of 1.5 ms were employed. The signal was recorded after applying a selective monitoring $\pi/2$ pulse to the ancilla ($C^\alpha$) and while applying a weak non-selective gradient (0.15 gauss/cm). The magnetic field gradient commutes with the diagonal states of the data spins so that it added a linear time dependent spatial phase exclusively to the ancilla. The time trace, given in Fig. 4, shows four gradient echoes corresponding to the four k-space encoded subspaces. The echoes are equally spaced in time, which implies on the evenly spaced discrete distribution of labeling in k-space.

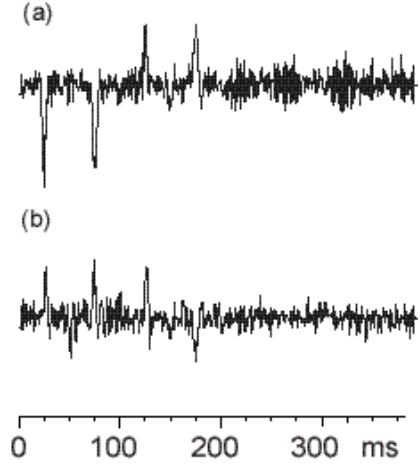

**Figure 4:** The real (a) and imaginary (b) parts of a time trace recorded immediately after a selective $(\pi/2)_y^a$ monitoring RF pulse and during a weak readout magnetic field gradient. It shows four echoes at $t_n = 0.025 \times (2n+1)$ where the echo numbers $n = 0, 1, 2,$ are associated with the subspaces $|10\rangle$, $|00\rangle$, $|01\rangle$ and $|11\rangle$, respectively. The four subspaces provided by two data spins were pre-encoded using the procedure in Fig. 2(b) with magnetic field gradient pulses chosen according to Eq. [18] (see text).

A readout gradient is a simple method to extract information from all subspaces, however, in addition to being non-selective it suffers from several drawbacks, namely, rapid loss of signal due to transverse relaxation ($T_2^*$), limited bandwidth, and poor spectral resolution. A better monitoring method involves a discrete scan of k-space. This is done using repeated pulses of magnetic field gradient so to satisfy the ramping condition Eq. (19). The method was implemented for the encoded state of two data spins and the spectra obtained for the four pseudo-pure states shown in Fig. 5. The fast switching rate (5 kHz) of the gradients, compared to the total spectral width of 600Hz, allowed acquisition of all four spectra in a single scan. The signal to noise ratio was somewhat lower than that obtained for the spectrum of a single subspace, mainly due molecular diffusion during the application of the gradients.

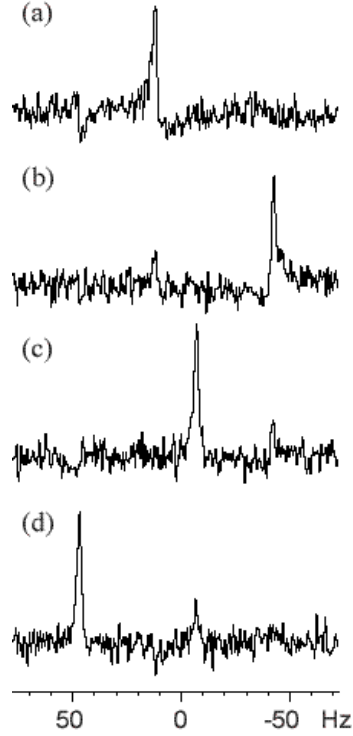

**Figure 5:** Fast decoding of all four pre-encoded subspaces: $|10\rangle$, $|00\rangle$, $|01\rangle$ and $|11\rangle$. (a)-(d): Spectra corresponding to the four states were obtained in a single scan through k-space by switching the applied magnetic field gradient rapidly according to the ramping condition $\Delta k = 2k_0$.

## VII. THE EFFECT OF MOLECULAR DIFFUSION

As implied earlier, molecular diffusion scrambles spatially encoded states in a process that leads to an irreversible loss of coherence. To quantify the importance of this process each wave number can be associated with a diffusive attenuation $A_k = e^{-\int k^2(t) D dt}$ according to the diffusion coefficient $D$ [16]. Two cases to consider are the attenuation during and after the encoding sequence. The overall attenuation during the preparation of a single pseudo-pure state can be calculated using Eq. (14) and the fact that $k_n = k_0$, yielding

$$A_N \times \cdots \times A_2 \times A_1 = e^{-N^2 k_0^2 (\Delta_N + \delta/3) D} \cdots e^{-2^2 k_0^2 (\Delta_2 + \delta/3) D} e^{-k_0^2 (\Delta_1 + \delta/3) D}. \qquad (23)$$

where $\Delta_n$ is the duration of the *n-th* c-NOT gate, and $\delta$ is the length of the of the magnetic field gradient pulse. The time required to perform each c-NOT operation varies due to the diversity in the J-couplings. In the simplified case where this time is assumed to be constant ($\Delta_n = \Delta$) the logarithm of the overall attenuation is $N(N+1)(2N+1)k_0^2(\Delta+\delta/3)D/6$. Similarly for the multi-encoding case, the overall attenuation can be calculated for each of the subspaces. The minimum and maximum attenuation occurs for the states $|0\rangle$ and the state $|2^N-1\rangle$, respectively. Using Eq. (11), Eq. (14) and Eq. (18) they are

$$A(|0\rangle) = e^{-(1-(-2)^N)^2 k_0^2 (\Delta_N+\delta/3)D} \cdots e^{-(-1)^2 k_0^2 (\Delta_2+\delta/3)D} e^{-k_0^2(\Delta_1+\delta/3)D},$$

$$A(|2^N-1\rangle) = e^{-(1-2^N)^2 k_0^2 (\Delta_N+\delta/3)D} \cdots e^{-(1-2^2)^2 k_0^2 (\Delta_2+\delta/3)D} e^{-k_0^2(\Delta_1+\delta/3)D}. \qquad (24)$$

The attenuation rate can be modulated by varying either the duration ($\delta$) of the gradient pulse or its unit strength $k_0$. In practice, this ability is restricted by the available resources such as dynamic range and switching time of the magnetic field gradients, as well as by sample characteristics (diffusion coefficient, size).

Once the encoding has been obtained, diffusion causes an exponential attenuation with a constant decay rate $(k_\alpha + k_s)D$. In the case of a single pseudo-pure state this decay rate ranges from $4k_0^2 D$ to $4N^2 k_0^2 D$. The slow decay rate dictates the time it takes molecular diffusion to decohere all $2^N-1$ "undesired" subspaces i.e. the terms in the sum of Eq. (15). For multi-encoding of $2^N$ pseudo-pure states, decay rate ranges from $k_0^2 D$ to $(2^N-1)^2 k_0^2 D$, and therefore, limits the available computation time if one wants to keep track of all pseudo-pure states. It does not however restrict the lifetime of a single pseudo-pure state of the form given in Eq. (16) that carries no spatial phase.

## VIII. DISCUSSION

k-space encoding provides a method for preparing an ensemble of spins, initially in thermal equilibrium, in a spatially conditional pseudo-pure state. The preparation of a single pseudo-pure state is done efficiently (with linear resources) and requires only one controlled-NOT gate for each data qubit. Unlike other methods the encoding step is projective and can be repeated leading to the same pseudo-pure state. The method is limited only by the initial polarization and relaxation times, and therefore in liquid-state NMR it can probably be extended to about 10 qubits.

The general scheme can be used to encode and decode more than one pseudo-pure state with only a little loss in signal intensity. Detection of multiple subspaces requires a fast switching time for the magnetic field gradient to provide sufficient bandwidth. In weakly coupled spin systems this requirement is determined by the distribution of coupling constants which is typically sparse. On the other hand, the maximum gradient strength is limited by decoherence which is dictated by molecular diffusion. Therefore, with typical diffusion and coupling constants, and currently available gradient systems, the method is probably restricted to few dozens of subspaces. Multi-encoding of pseudo-pure states could be useful in various areas such as quantum tomography, the study of decoherence and simulations of a quantum ensemble.

The idea of a conditional spatial phase shift can be extended to other conditional logic gates. In general one can define an operator $D_k^{a|w}$ that will shift the phase of the ancilla conditional on any state $|w\rangle$. For example, the condition $|w\rangle = |11\rangle$ leads to a Toffoli type of phase shift $D_{k_1 k_2}^{a|11}$, i.e. it adds a spatial phase to the ancilla conditional on both spin 1 and spin 2 being in the state $|1\rangle$. With a proper choice of magnetic field gradients, the application of such an operator to the state $\sigma_x^a$ would result in a two-spin pseudo-pure state $\sigma_x^a E_-^1 E_-^2$. Alternatively, it can be used not as a method of encoding pseudo-pure states, but later on during a computation to permute the k-labels conditional on the state of the system, perhaps labeling the desired result in this fashion.

Molecular diffusion adds an inhomogeneous decay mechanism whose strength is experimentally controllable. Combined with conditionally spatial phase shifts, molecular diffusion offers a general method for implementing projections of an arbitrary state onto a pre-selected subspace, and thereby mimicking projective measurements. Finally, the ability of the method to decohere selected subspaces while retaining the information in others further extends the class of spin dynamics and decoherence processes that can be emulated by liquid-state NMR.

## ACKNOWLEDGMENTS

This work was supported by the U.S. Army Research Office under grant number DAAG-55-97-1-0342 from the DARPA Microsystems Technology Office.